\documentclass[a4paper,12pt]{article}
\usepackage{graphicx}
\usepackage{amsmath,amssymb,amsthm}

\newtheorem{theorem}{Theorem}[section]

\newtheorem{proposition}[theorem]{Proposition}
\newtheorem{lemma}[theorem]{Lemma}

\newtheorem{definition}[theorem]{Definition}

\numberwithin{equation}{section} \numberwithin{theorem}{section}
\newcommand{\prob}{{\mathbb P}}
\newcommand{\expec}{{\mathbb E}}

\title{Uniqueness of the stationary distribution and stabilizability 
in Zhang's sandpile model}
\author{Anne Fey-den Boer\footnote{TU Delft, The Netherlands, a.c.fey.denboer@tudelft.nl} \and
Haiyan Liu\footnote{VU University Amsterdam,
De Boelelaan 1081a, 1081 HV Amsterdam, The Netherlands,
hliu@few.vu.nl} \and Ronald Meester\footnote{VU University Amsterdam,
De Boelelaan 1081a, 1081 HV Amsterdam, The Netherlands, rmeester@few.vu.nl}}
\begin{document}
\maketitle

\begin{abstract}
We show that Zhang's sandpile model $(N, [a, b])$ on $N$ sites and with
uniform additions on $[a,b]$ has a unique
stationary measure for all $0\leq a < b\leq 1$. This generalizes
earlier results of \cite{anne} where this was shown in some special cases.

We define the infinite volume Zhang's sandpile model in dimension
$d\geq1$, in which topplings occur according to a Markov toppling
process, and we study the stabilizability of initial configurations
chosen according to some measure $\mu$. We show that for a
stationary ergodic measure $\mu$ with density $\rho$, for all
$\rho<\frac{1}{2}$, $\mu$ is stabilizable; for all $\rho\geq 1$,
$\mu$ is not stabilizable; for $\frac{1}{2}\leq \rho<1$, when
$\rho$ is near to $\frac{1}{2}$ or $1$, both possibilities can occur.
\end{abstract}

\section{Introduction}
Zhang's sandpile model \cite{zhang} is a variant of the more
common abelian sandpile model \cite{dhar}, which was introduced in
\cite{BTW} as a toy model to study self-organized criticality. 
We define the model more precisely in the next section, but we start
here with an informal discussion. 

Zhang's model differs from the abelian sandpile model on a finite
grid $\Lambda$ in the following respects: The configuration space
is $[0,1)^\Lambda$, rather than $\{0,1,\ldots,2d-1\}^\Lambda$. The model evolves, like
the abelian sandpile model, in discrete time through additions and
subsequent stabilization through topplings of unstable sites.
However, in Zhang's model, an addition consists of a continuous
amount, uniformly distributed on $[a,b] \subseteq
[0,1]$, rather than one `grain'. Furthermore, in a Zhang toppling of
an unstable site, the entire height of this site is distributed
equally among the neighbors, whereas in the abelian sandpile model
one grain moves to each neighbor irrespective of the height of the
toppling site.

Since the result of a toppling depends on the height of the
toppling site, Zhang's model is not abelian. This means that
`stabilization through topplings' is not immediately well-defined. However,
as pointed out in \cite{anne}, when we work on the line, and as long as there are no two neighbouring
unstable sites, topplings are abelian. When the initial configuration 
is stable, we will only encounter realizations with no two neighbouring 
unstable sites, and we have - a fortiori - that the model is abelian. 

In \cite{anne}, the following main results, in dimension 1, were obtained.
Uniqueness of the stationary measure was proved for a number
of special cases: (1) $a\geq \frac{1}{2}$; (2) 
$N=1$, and (3) $[a,b] = [0,1]$. For the model on one
site with $a=0$, an explicit expression for the stationary height
distribution was obtained. Furthermore, the existence of so called
`quasi-units' was proved for $a \geq 1/2$, that is, in the
limit of the number of sites to infinity, the one-dimensional marginal of the stationary
distribution concentrates on a single value $\frac{a+b}2$.

In the first part of the present paper, we prove, in dimension 1,  uniqueness of the
stationary measure for the general model, via a coupling which is much more
complicated than the one used in \cite{anne} for the special case $a  \geq 1/2$.

In Section \ref{densitysection}, we study an infinite-volume
version of Zhang's model in {\em any} dimension. A similar infinite-volume version of
the abelian sandpile model has been studied in
\cite{quant,feyredig} and we will in fact also use some of the ideas in that paper.

For Zhang's infinite-volume model in dimension $\geq 1$, we start
from a random initial configuration in
$[0,\infty)^{\mathbb{Z}^d}$, and evolve it in time by Zhang
topplings of unstable sites. We are
interested in whether or not there exists a limiting stable configuration.
Since Zhang topplings are not
abelian, for a given configuration $\eta\in[0,
\infty)^{\mathbb{Z}^d}$, for some sequence of topplings it may
converge to a stable configuration but for others, it may not. Moreover
we do not expect the final configuration - if there is any - to be unique. Therefore, we choose a random order of topplings as follows. To every site we attach an independent rate 1 Poisson clock, and
when the clock rings, we topple this site if it is unstable; if it is stable we do nothing. For
obvious reasons we call this the {\em Markov toppling process}.

We show that if we choose the initial configuration according to a
stationary ergodic measure $\mu$ with density $\rho$, then for all
$\rho<\frac{1}{2}$, $\mu$ is stabilizable, that is, the configuration converges to a
final stable configuration. For all $\rho\geq 1$,
$\mu$ is not stabilizable. For $\frac{1}{2}\leq \rho<1$, when
$\rho$ is near to $\frac{1}{2}$ or $1$, both possibilities can occur.

\section{Model definition and notation}

In this section, we discuss Zhang's sandpile on $N$ sites, labelled $1,2,\ldots,N$. We denoted by
$\mathcal{X}_{N}=[0, \infty)^{N}$ the set of all possible
configurations in Zhang's sandpile model. We will use symbols
$\eta$ and $\xi$ to denote a configuration. We denote the value of
a configuration $\eta$ at site $x$ by $\eta_{x}$, and refer to
this value as the {\em height}, {\em mass} or {\em energy} at site $x$. 
We introduce a labelling of sites
according to their height, as follows.

\noindent
\begin{definition}
For $\eta\in\mathcal{X}_{N}$, we call site $x$
$$
\begin{array}{lll}
\textrm{empty} & \textrm{if} & \eta_x = 0,\\
\textrm{anomalous} & \textrm{if} & \eta_x \in (0,\frac 12),\\
\textrm{full} & \textrm{if} & \eta_x \in [\frac 12,1),\\
\textrm{unstable} & \textrm{if} & \eta_x \geq 1.\\
\end{array}
$$
\end{definition}
A site $x$ is stable for $\eta$ if $0\leq\eta_x<1$, and hence all the empty,
anomalous and full sites are stable. A configuration
$\eta$ is called stable if all sites are stable, otherwise
$\eta$ is unstable. $\Omega_N = [0,1)^N $ denotes the set of all
stable configurations.

By $T_x(\eta)$ we denote the (Zhang) toppling operator for site $x$,
acting on $\eta$ and which is defined as follows.

\begin{definition}
For all $\eta\in\mathcal{X}_{N}$ such that $\eta_{x} \geq
1$, we define
\begin{equation*}
\label{top} T_x (\eta)_{y} = \left\{
\begin{array}{ll}
0 & \mbox{ if } x=y,\\
\eta_{y} + \frac{1}{2}\eta_{x} & \mbox{ if } |y-x|=1,\\
\eta_{y} & \mbox{ otherwise}.
\end{array}
\right.
\end{equation*}
For all $\eta$ such that $\eta_{x}<1$,$ T_x(\eta) = \eta $, for
all $x$.
\end{definition}

In other words, the toppling operator only changes $\eta$ if site
$x$ is unstable; in that case, it divides its energy in equal
portions among its neighbors. We say in that case that site $x$
{\em topples}. If a boundary site topples, then half of its energy
disappears from the system.
Every configuration in $\mathcal{X}_N$ can stabilize, that is,
reach a final configuration in $\Omega_N$, through finitely many
topplings of unstable sites, since energy is dissipated at the
boundary. 

We define the $(N,[a,b])$ model as a discrete time Markov process
with state space $\Omega_N$, as follows. The process starts at
time 0 from configuration $\eta(0)=\eta$. For every $t =
1,2,\ldots$, the configuration $\eta(t)$ is obtained from
$\eta(t-1)$ as follows: a random amount of energy $U(t)$,
uniformly distributed on $[a,b]$, is added to a uniformly chosen
random site $X(t)\in \{ 1,\ldots, N\}$, hence $\mathbb{P}(X(t) =
j)=1/N$ for all $j=1,\ldots, N$. The random variables $U(t)$ and $X(t)$ 
are independent
of each other and of the past of the process. We stabilize the
resulting configuration through topplings (if it is already
stable, then we do not change it), to obtain the new configuration
$\eta(t)$. By $\expec^\eta$ and $\prob^\eta$, we denote
expectation resp. probability with respect to this process.

\section{Uniqueness of the stationary distribution}
\label{generalcouplingsection}

In Zhang's sandpile model, it is not obvious that the stationary
distribution is unique, since the state space is uncountable. For
the three cases: (1) $N=1$, (2) $a\geq \frac{1}{2}$ and $N \geq
2$, (3) $a=0, b=1$ and $N \geq 2$, it is shown in \cite{anne} that
the model has a unique stationary distribution, and in addition,
in case (2) and (3), for every initial distribution $\nu$, the
measure at time $t$, denoted by $\nu_{t}^{a, b, N}$,  converges in
total variation to the stationary distribution. In the case $N=1$,
there are values of $a$ and $b$ where we only have time-average
total variation convergence, see Theorem 4.1 of \cite{anne}.

In all these cases (except when $N=1$) the proof consisted of constructing a coupling
of two copies of Zhang's model with arbitrary initial
configurations, in such a way that after some (random) time, the
two coupled processes are identical. We will call such a coupling
`successful', as in  \cite{thorisson}. Each coupling was very
specific for the case considered. In the proof for the case
$a\geq\frac{1}{2}$ and $N \geq 2$, explicit use is being made of
the fact that an addition to a full site always causes a toppling.
The proof given for the case $a=0$ and $b=1$ and $N>1$ can be
generalized to other values of $b$, but $a=0$ is necessary, since
in the coupling we use that additions can be arbitrarily small. In
the special case $N=1$, the model is a renewal process, and the
proof relies on that.
Here is the main result of this section; note that only
the case $a=b$ is not included.
 
\begin{theorem} For every $0\leq{a}<b\leq1$,
and $N\geq2$, Zhang's sandpile model $(N, [a,b])$ has a unique stationary distribution 
which we denote by $\mu^{a, b,N}$.
Moreover, for every initial distribution on $[0, 1)^N$,
the distribution of the process at time $t$ converges in total variation 
to $\mu^{a, b,N}$, as $t \to \infty$. 
\label{generalcasetheorem}
\end{theorem}

We introduce some notation. Denote $\eta,\xi\in\Omega_N$ as the
initial configurations, and  $\eta(t), \xi(t)$ as two independent
copies of the processes, starting from $\eta$ and $\xi$ respectively. 
The independent additions at time $t$ for
the two processes starting from $\eta, \xi$ are $U^{\eta}(t)$ and
$U^{\xi}(t)$, addition sites are $X^{\eta}(t), X^{\xi}(t)$
respectively. 
Often, we will use `hat'-versions of the various quantities to denote
a coupling between two processes. So, for instance, $\hat{\eta}(t), \hat{\xi}(t)$ denote coupled
processes (to be made precise below) with initial configurations $\eta$ and $\xi$
respectively. By $\hat{X}^\eta(t)$ and $\hat{X}^\xi(t)$ we denote
the addition sites at time $t$ in the coupling, and by
$\hat{U}^\eta(t)$ and $\hat{U}^\xi(t)$ the addition amounts at
time $t$.

In this section, we will encounter configurations that are such
that they are empty at some site $x$, $1\leq x \leq N$, and full
at all the other sites. We denote the set of such configurations
$\mathcal{E}_x$. By $\mathcal{E}_{b}$, we denote the set of
configurations that have only one empty boundary site, and are
full at all other sites, that is,
$\mathcal{E}_{b}=\mathcal{E}_{1}\cup \mathcal{E}_{N}$.

In order to give the proof of Theorem \ref{generalcasetheorem}, we
need the following three preliminary results, the proof of which
will be given in Sections \ref{step1subsection},
\ref{step2subsection} and \ref{step3subsection} respectively.

\begin{lemma}
For all $\eta$, $\xi$ in $\Omega_{N}$, $\eta(t)$ and $\xi(t)$ are a.s.\
simultaneously in $\mathcal{E}_b$ infinitely often.
\label{step1lemma}
\end{lemma}

\begin{lemma}
Let $\eta$ and $\xi$ be two configurations in $\mathcal{E}_N$ and let, for 
all $\epsilon >0$, 
$$
t_\epsilon =  2\lceil
\frac{2}{a+b} \rceil \cdot\ \lceil \log_{(1 -2^{-\lceil \frac{3N}2
\rceil})}(\frac {2\epsilon}N)\rceil.
$$
Consider couplings  $(\hat{\eta}(t), \hat{\xi}(t))$ of the
process starting at $\eta$ and $\xi$ respectively.
Let, in such a coupling, $T$ be the first time $t$ with the property that 
\begin{equation}
\max_{1\leq x\leq
N}\mid\hat{\eta}_x(t)-\hat{\xi}_x(t)\mid<\epsilon \label{req0}
\end{equation}
and
\begin{equation}
\hat{\eta}(t)\in\mathcal{E}_N, \hat{\xi}(t)\in\mathcal{E}_N.
\label{req1}
\end{equation}
There exists a coupling such that the event $T \leq t_{\epsilon}$ 
has probability at least $(2N)^{-t_\epsilon}$, uniformly in $\eta$ and $\xi$.
\label{step2lemma}
\end{lemma}

\begin{lemma}
Let 
$$
\epsilon_{a,b,N}= \frac{b-a}{6+16\Pi_{l=1}^{N-1}(1+2^{N-2-l})}.
$$
Consider couplings  $(\hat{\eta}(t), \hat{\xi}(t))$ of the
process starting at $\eta$ and $\xi$ respectively, with the property that
\begin{equation}
\max_{1\leq x\leq N}
\mid\eta_{x}-\xi_{x}\mid<\epsilon_{a,b,N}. 
\label{req2}
\end{equation}
Let $T'$ be the first time $t$ with the property that $\hat{\eta}(t)=\hat{\xi}(t)$.
Then there exists a coupling such that the event 
$T' < (N-1)\lceil\frac{1}{a+b}\rceil$ 
has probability bounded below by a positive constant that depends only on $a$, $b$ and $N$.
\label{step3lemma}
\end{lemma}

We now present the coupling that constitutes the proof of
Theorem \ref{generalcasetheorem}, making use of the results stated
in Lemma \ref{step1lemma}, Lemma \ref{step2lemma} and Lemma
\ref{step3lemma}.

\medskip\noindent {\it Proof of Theorem \ref{generalcasetheorem}.} Take two
probability distributions $\nu_{1}, \nu_{2}$ on $\Omega_N$, and
choose $\eta$ and $\xi$ according to $\nu_{1}, \nu_{2}$
respectively. We present a successful coupling
$\{\hat{\eta}(t),\hat{\xi}(t)\}$, with $\hat{\eta}(0)=\eta$ and
$\hat{\xi}(0)=\xi$. If we assume that both $\nu_{1}$ and $\nu_{2}$
are stationary, then the existence of the coupling shows that
$\nu_1 = \nu_2 = \nu$. If we take $\nu_1=\nu$ and $\nu_2$
arbitrary, then the existence of the coupling shows that any
initial distribution $\nu_2$ converges in total variation to
$\nu$.

The coupling consists of three steps, and is described
as follows.

\begin{itemize}
\item {\em step 1.} We evolve the two processes independently until
they encounter a configuration in $\mathcal{E}_b$ simultaneously.
From Lemma \ref{step1lemma} we know this happens a.s.
By symmetry, we can assume that both configurations are in $\mathcal{E}_N$ as soon as both
processes have reached a configuration in $\mathcal{E}_b$. From
that moment on, we proceed to 

\item {\em step 2.} We use the coupling
as described in the proof of Lemma \ref{step2lemma}. That amounts
to choosing $\hat{X}^{\xi}(t) = \hat{X}^{\eta}(t) = X^\eta(t)$,
and  $\hat{U}^{\xi}(t) = \hat{U}^{\eta}(t) = U^\eta(t)$. As the
proof of Lemma \ref{step2lemma} shows, if $U^\eta(t)$ and
$X^\eta(t)$ satisfy certain requirements for at most
$t_{\epsilon}$ time steps, then we have that \eqref{req0}
and \eqref{req1} occur, with $\epsilon = \epsilon_{a,b,N}$. If
during this step, at any time step either $U^\eta(t)$ or
$X^\eta(t)$ does not satisfy the requirements, then we return to
step 1. But once we have \eqref{req0} and \eqref{req1} (which, by
Lemma \ref{step2lemma}, has positive probability), then we proceed
to

\item {\em step 3.} Here, we use the coupling as described in the
proof of Lemma \ref{step3lemma}. Again, we choose
$\hat{X}^{\xi}(t) = \hat{X}^{\eta}(t) = X^\eta(t)$ and
$\hat{U}^{\eta}(t) = U^\eta(t)$, but the dependence of
$\hat{U}^{\xi}(t)$ on $U^\eta(t)$ is more complicated; the details can
be found in the proof of Lemma \ref{step3lemma}. As the proof of
Lemma \ref{step3lemma} shows, if $U^\eta(t)$ and $X^\eta(t)$
satisfy certain requirements for at most $(N-1)\lceil \frac 1{a+b}
\rceil$ time steps, then we have that $\hat{\eta}(t)=
\hat{\xi}(t)$ occurs, and from that moment on the two
processes evolve identically. By Lemma \ref{step3lemma}, this event has
positive probability. If during this step, at any time
step, either $U^\eta(t)$ or $X^\eta(t)$ does not satisfy the
requirements, we return  to step 1.
\end{itemize}

In the coupling, we keep returning to step 1 until step 2 and
subsequently step 3 are successfully completed, after which we
have that $\hat{\eta}(t)= \hat{\xi}(t)$. Since each step is successfully completed with
uniform positive probability, we a.s.\ need only finitely many attempts.
Therefore we achieve $\hat{\eta}(t)= \hat{\xi}(t)$ in finite time,
so that the coupling is successful. 
\qed

\medskip
Now, we will proceed to give the proof of Lemma \ref{step1lemma},
Lemma \ref{step2lemma} and Lemma \ref{step3lemma}.

\subsection{Proof of Lemma \ref{step1lemma}}
\label{step1subsection} 
In this section, we show that starting two independent processes
from any two configurations $\eta$ and $\xi$, the two processes
will a.s.\ be in $\mathcal{E}_b$ simultaneously infinitely often. The
proof will be realized in two steps.

\begin{lemma}
Let $\eta$ be a configuration in $\Omega_N$. The process starting from 
$\eta$ visits
$\mathcal{E}_N$ within $(N+1)\lceil\frac{1}{a+b}\rceil$ time steps,
with probability at least $(\frac{1}{2N})^{(N+1)\lceil
\frac{1}{a+b} \rceil}$. \label{EN1lemma}
\end{lemma}

\begin{proof}
We prove this by giving an explicit event realizing this, that has
the mentioned probability. In this step, we always make {\em heavy}
additions to site $N$, that is, additions with value at least $(a+b)/2$.

First, starting from configuration $\eta$,  
we make heavy additions to site $1$ until site $1$ becomes
unstable. Then an avalanche occurs and a new configuration with
at least one empty site is reached. The leftmost empty site denoted
by $r_1$. If $r_1=N$ we are done. If $r_1 \neq N$, then it is easy to check that site $r_1+1$ is full. The total number of additions
needed for this step is at most $2\lceil\frac{1}{a+b}\rceil$.

Then, if $r_1\neq N$, we continue by making heavy
additions to site $r_1+1$ until site $r_1+1$ becomes unstable. Then an
avalanche starts from site $r_1+1$. During this avalanche, sites $1$ to $r_1-1$ are not affected, site $r_1$ becomes full and we again reach a new configuration with at least one empty site, the leftmost of which is denoted by $r_2$. If $r_2=N$ we are done. If not, note that $r_2\geq r_1+1$ and that all sites $1, ..., r_2-1$ and 
$r_2+1$ are full. At most $\lceil\frac{1}{a+b}\rceil$ heavy additions are needed for this step.

If $r_2\neq N$, we repeat this last procedure. After each
avalanche, the leftmost empty site moves at least one site to the right, and hence,
after the first step we need at most $N-1$ further steps. 

Hence, the total number of heavy additions needed for the above
steps is bounded above by $(N+1)\lceil\frac{1}{a+b}\rceil$. Every
time step, with probability $\frac{1}{N}$, a fixed site is chosen and
with probability $\frac{1}{2}$, an addition is a heavy addition. Therefore,
the probability of this event is at least $(2N)^{-(N+1)\lceil
\frac{1}{a+b}\rceil}$.
\end{proof}

\begin{lemma}
Let $\xi(0)\in\mathcal{E}_{b}$, then $\xi(1)\in\mathcal{E}_b$ with
probability at least $\frac 1N$. \label{EN2lemma}
\end{lemma}

\begin{proof}
Again, we give an explicit possibility with probability $\frac
1N$. Starting in $\xi \in \mathcal{E}_b$, we make one addition to
the site next to the empty boundary site. If this site does not
topple, then of course $\xi(1)$ is still in $\mathcal{E}_b$. But
if it does topple, then every full site will topple once, after
which all sites will be full except for the opposite (previously
full) boundary site. In other words, then $\xi(1)$ is also in
$\mathcal{E}_b$. The probability that the addition site is the
site next to the empty boundary site, is $\frac 1N$. Then
$\xi(1)\in\mathcal{E}_b$ with probability at least $\frac 1N$.
\end{proof}

\medskip\noindent{\em Proof of Lemma \ref{step1lemma}}. 
From Lemma \ref{EN1lemma}, it follows that the process starting from $\xi$ is in
$\mathcal{E}_b$ infinitely often. Let $t^{\xi}_{b}$ be the first time that the process is in $\mathcal{E}_b$, and define
$$
T^{\eta}_b=\min\{t:t\geq t_b^\xi,\eta(t) \in \mathcal{E}_b\}.
$$
By the same lemma, the probability that $0\leq
T^{\eta}_b-t^{\xi}_b \leq (N+1)\lceil\frac{1}{a+b}\rceil$ is at
least $(2N)^{-(N+1)\lceil\frac{1}{a+b}\rceil}$.

Repeatedly applying Lemma \ref{EN2lemma} gives that the event that
$\xi(t^{\xi}_b+1)\in\mathcal{E}_b,
\xi(t^{\xi}_b+2)\in\mathcal{E}_b, ...,
\xi(T^{\eta}_b)\in\mathcal{E}_b$, occurs with probability bounded
below by $(\frac{1}{N})^{T^{\eta}_b-t^{\xi}_b}$.

We have showed that when $\xi(t) \in
\mathcal{E}_b$, within at most $(N+1)\lceil\frac{1}{a+b}\rceil$ time
steps, the two processes are in $\mathcal{E}_b$ simultaneously
with probability at least
$(\frac{1}{2N^2})^{(N+1)\lceil\frac{1}{a+b}\rceil}$. Combining
this with the fact that the process starting from $\xi$ is in
$\mathcal{E}_b$ infinitely often, we conclude the two processes are in
$\mathcal{E}_{b}$ simultaneously infinitely often.
\hfill $\Box$\\

\subsection{Proof of Lemma \ref{step2lemma}}
\label{step2subsection} In this part, we couple two processes
starting from $\eta,\xi \in \mathcal{E}_N$. The coupling consists
of choosing the addition amounts and sites equal at each time
step. For this coupling, we present an event that has probability
$(2N)^{-t_\epsilon}$, with $t_\epsilon = 2\lceil \frac{2}{a+b}
\rceil \cdot\ \lceil \log_{(1 -2^{-\lceil \frac{3N}2
\rceil})}(\frac {2\epsilon}N)\rceil$, and which is such that if
it occurs, then \eqref{req0} and \eqref{req1} are satisfied.

The event we need is that for $t_\epsilon$ time steps,
\begin{enumerate}
\item All additions are heavy. 
\item The additions occur
to site $N$ until site $N$ becomes unstable, then to site $1$
until site $1$ becomes unstable, then to site $N$ again, etcetera.
\end{enumerate}
The probability for an addition to be heavy is $\frac 12$ and the
probability for the addition to occur to a fixed site is $\frac
1N$. Therefore, the probability of this event is
$(2N)^{-t_\epsilon}$.

Now we show that if this event occurs, then \eqref{req0} and
\eqref{req1} are satisfied.
Let $\hat{U}(t)$ be the addition amount at time $t$. Define a
series of stopping times $\{\tau_{k}\}_{k\geq0}$ by
\begin{equation}
\tau_{0}=0, \tau_{k}: =\min\{t>\tau_{k-1}:
\sum_{t=\tau_{k-1}+1}^{\tau_{k}}\hat{U}(t)\geq1\}, \textrm{ for }
k\geq1,
\end{equation}
and denote
\begin{equation}
S_{k}=\sum_{t=\tau_{k-1}+1}^{\tau_{k}}\hat{U}(t).
\end{equation}

The times $\tau_k$ ($k > 0$) are such that in both configurations,
{\em only} at these times an avalanche occurs. Indeed, for the first avalanche
this is clear because we only added to site $N$, which was empty
before we started adding. But whenever an avalanche starts at a
boundary site, and all other sites are full, then every site
topples exactly once and after the avalanche, the opposite
boundary site is empty. Thus the argument applies to all
avalanches.

Since we make only heavy additions,
\begin{equation}
\tau_k-\tau_{k-1}\leq\lceil\frac 2{a+b}\rceil, \textrm{ for all }
k. \label{timedifference}
\end{equation}

After the $k$-th avalanche, the height $\hat{\eta}_y(\tau_k)$ is a
linear combination of $S_1, ..., S_k$ and $\eta_1, ...,
\eta_{N-1}$, which we write as

\begin{eqnarray}
\hat{\eta}_{y}(\tau_{k}) & = &
\sum_{l=1}^{k}A_{ly}(k)S_{l}+\sum_{m=1}^{N}B_{my}(k){\eta}_{m},
\textrm{ for } 1 \leq {y} \leq {N},
 \label{newconfiguration}
\end{eqnarray}
and a similar expression for $\hat{\xi}_{y}(\tau_{k})$.
From Proposition 3.7 of \cite{anne}, we have that
$$
B_{my}(k)\leq (1-2^{-\lceil\frac{3N}{2}\rceil})\max_{x}B_{mx}(k-1).
$$
By induction, we find
$$
B_{my}(k)\leq (1-2^{-\lceil\frac{3N}{2}\rceil})^{k}
$$
and hence
\begin{eqnarray*}
\max_{1\leq y\leq
N}\mid\hat{\eta}_{y}(\tau_{k})-\hat{\xi}_{y}(\tau_{k})\mid &\leq&
N(1-2^{-\lceil\frac{3N}{2}\rceil})^{k} \max_{1\leq x\leq
N}\mid\eta_x-\xi_x\mid\\
&\leq& \frac{N}{2}(1-2^{-\lceil\frac{3N}{2}\rceil})^{k},
\end{eqnarray*}
where we use the fact that $\eta, \xi \in\mathcal{E}_N$ implies 
$\max_{1\leq x \leq
N}\mid\eta_x-\xi_x\mid\leq\frac{1}{2}$.

For each $\epsilon>0$, choose $k_\epsilon= 2\lceil \log_{ (1
-2^{-\lceil \frac{3N}2 \rceil})}(\frac{2\epsilon}N) \rceil$. Then
$\frac{N}{2}(1-2^{-\lceil\frac{3N}{2}\rceil})^{k_{\epsilon}}\leq
\epsilon$, so that
\begin{equation*}
\max_{1\leq x \leq N}\mid{\hat{\eta}}_{x}
(\tau_{k_{\epsilon}})-\hat{\xi}_{x}(\tau_{k_{\epsilon}})\mid < \epsilon,
\end{equation*}
and moreover, an even number of avalanches occurred, which means
that at time $\tau_{k_\epsilon}$, both processes are in
$\mathcal{E}_{N}$. By \eqref{timedifference}, $\tau_{k_\epsilon}
\leq t_\epsilon = k_\epsilon \lceil \frac 2{a+b} \rceil$. Thus,
$\tau_{k_\epsilon}$ is a random time $T$ as in the statement of Lemma
\ref{step2lemma}.
\qed

\subsection{Proof of Lemma \ref{step3lemma}}
\label{step3subsection}

As in the proof of Lemma \ref{step2lemma}, we will describe the
coupling, along with an event that has probability bounded below
by a constant that only depends on $a$, $b$ and $N$, and is such
that if it occurs, then within $(N-1)\lceil\frac{1}{a+b}\rceil$
time steps, $\hat{\eta}(t)=\hat{\xi}(t)$.
First we explain the idea behind the coupling and this event, after that
we will work out the mathematical details.

The idea is that in both processes we add the same amount, only to
site 1, until an avalanche is about to occur. We then add slightly
different amounts while still ensuring that an avalanche occurs in
both processes. After the first avalanche, all sites contain a
linear combination of the energy before the last addition, plus a
nonzero amount of the last addition. We choose the difference
$D_1$ between the additions that cause the first avalanche, such
that site $N$ will have the same energy in both processes after
the first avalanche, where $|D_1|$ is bounded above by a value
that only depends on $a$, $b$ and $N$. Sites $N-1$ will become
empty, and the differences between the two new configurations on
all other sites will be larger than those before this avalanche,
but can be controlled.

When we keep adding to site 1, in the next avalanche only the
sites $1, \ldots, N-2$ will topple. We choose the addition amounts
such that after the second avalanche, sites $N-1$ will have the
same energy. Since site $N$ did not change in this avalanche, we
now have equality on two sites. After the second avalanche, site
$N-2$ is empty, and the configurations are still more different on
all other sites, but the difference can again be controlled.

We keep adding to site 1 until, after a total of $N-1$ avalanches,
the configurations are equal on all sites. After each avalanche,
we have equality on one more site, and the difference increases on
the nonequal sites. We deal with this increasing difference by
controlling the maximal difference between the corresponding sites
of the two starting configurations by the constant $\epsilon_{a, b,
N}$, so that we can choose each
addition of both sequences from a nonempty interval in $[a,b]$.
The whole event takes place in finite time, and will therefore have
positive probability.

\medskip\noindent{\em Proof of Lemma \ref{step3lemma}}.
The coupling is as follows. We choose $\hat{X}^\eta(t) =
X^\eta(t)$, and $\hat{U}^\eta(t) = U^\eta(t)$. We choose the
addition sites $\hat{X}^\eta(t)$ and $\hat{X}^\xi(t)$ equal, and
the addition amounts $\hat{U}^\eta(t)$ and $\hat{U}^\xi(t)$ either
equal, or not equal but dependent. In the last case,
$\hat{U}^\xi(t)$ is always of the form $a + (\hat{U}^\eta(t) + D -
a) \mod (b-a)$, where $D$ does not depend on $\hat{U}^\eta(t)$.
The reader can check that, for any $D$, $\hat{U}^\xi(t)$ is then
uniformly distributed on $[a,b]$.

The event we need is as follows. First, all additions are heavy. For the
duration of $N-1$ avalanches, which is at most
$(N-1)\lceil\frac{1}{a+b}\rceil$ time steps, the additions to
$\eta$ occur to site 1, and the amount is for every time
step in a certain subinterval of $[a,b]$, to be specified next.
 
We denote $\frac{a+b}2 =
a''$ and recursively define
$$
 \epsilon_{k+1} = (1+2^{N-k-2}) \epsilon_k,
$$
with $\epsilon_1 = \epsilon_{a,b,N}$.
Between the $(k-1)$-st and $k$-th avalanche, the
interval is $[a'', a''+2\epsilon_k]$, and at the time where the
$k$-th avalanche occurs, the interval is a subinterval of
$[a',b] = [a''+3\epsilon_k, b]$ of length $\frac{b-a'}2$; see below.

The probability of at most $(N-1)\lceil\frac{1}{a+b}\rceil$
additions occurring to site 1, is bounded from below by
$N^{-\lceil \frac 1{a+b} \rceil (N-1)}$. Since $\epsilon_k$ is
increasing in $k$, the probability of all addition amounts
occurring in the appropriate intervals, is bounded below by
$(2\epsilon_1)^{(N-1)(\lceil \frac 1{a+b} \rceil -1)} \cdot
(\frac{b-a}4 - \frac {3\epsilon_{(N-1)}}2 )^{N-1}$. The
probability of the event is bounded below by the product of
these two bounds.

Now we define the coupling such that if this event occurs, then
after $N-1$ avalanches, we have that $\hat{\eta}(t)=\hat{\xi}(t)$.
In the remainder, we suppose that the above event occurs.

We start with discussing the time steps until the first avalanche.
Suppose, without loss of generality, that $\eta_1\geq\xi_1$. We
make equal additions in $[a'', a''+2\epsilon_1]$, until the first
moment $t$ such that $\eta_{1}(t) > 1-a''-2\epsilon_1$. We then
know that $\xi_1(t) > 1-a''-3\epsilon_1$. If we now choose the
last addition for both configurations in $[a',b]=[a''+3\epsilon_1,
b]$, then both will topple. 

Define
\begin{equation}
D_1:=\sum_{y=1}^{N-1}2^{y-1}(\eta_y-\xi_y). 
\label{Ueis}
\end{equation}

Let $\tau_1$ be the time at which the first avalanche occurs. Then
we choose for all $t < \tau_1$,
$\hat{U}^{\eta}(t)=\hat{U}^{\xi}(t)$, and
$\hat{U}^{\xi}(\tau_1)-\hat{U}^{\eta}(\tau_1)=D_1$.
Since $|D_1| < 2^{N-1}\epsilon_1 \leq \frac{b-a'}{4}$, when
$\hat{U}^{\eta}(\tau_1) \in [\frac{3a'+b}4, \frac {a'+3b}4]$ (the
middle half of $[a',b]$)
\begin{equation}
a' \leq \hat{U}^{\eta}(\tau_{1})+D_{1}<b.
\end{equation}
So, if  $\hat{U}^{\eta}(\tau_1)$ is uniformly distributed on
$[\frac{3a'+b}4, \frac {a'+3b}4]$, then $\hat{U}^{\xi}(\tau_{1}) =
\hat{U}^{\eta}(\tau_1) + D_1$ is uniformly distributed on
$[\frac{3a'+b}4+D_{1}, \frac {a'+3b}4+D_{1}] \subset [a',b]$.

Let $R_1 = \sum_{t=1}^{\tau_1-1} \hat{U}^{\eta}(t)$. Then at time
$\tau_1$, for $1 \leq x \leq N-2$ we have
$$
\hat{\eta}_{x}(\tau_{1}) = \frac{1}{2^{x+1}}(\eta_1+R_{1}+\hat{U}^{\eta}(\tau_1))+\frac{1}{2^{x}}\eta_2+\cdots+\frac{1}{2}\eta_{x+1},
$$
and
$$
\hat{\eta}_{N-1}(\tau_1) = 0; 
\hat{\eta}_{N}(\tau_1) = \hat{\eta}_{N-2}(\tau_{1}),
$$
and a similar expression for $\hat{\xi}_x(\tau_1)$. It follows
that
$$
\hat{\eta}_{N}(\tau_1) -
\hat{\xi}_{N}(\tau_1)=-2^{(1-N)}D_1+\sum_{y=1}^{N-1}2^{y-N}(\eta_y-\xi_y)=0
$$
\noindent which means that the two coupled processes at time
$\tau_{1}$ are equal at site $N$.

After this first avalanche, the differences on sites $1,\ldots,
N-3$ have been increased. Ignoring the fact that sites $N-2$ happen to
be equal (to simplify the discussion), we calculate 
{\setlength\arraycolsep{-62.5pt}
\begin{eqnarray}
\max_{1\leq x \leq
N}\mid\hat{\eta}_{x}(\tau_1)-\hat{\xi}_{x}(\tau_1)\mid \leq
\max_{1\leq x \leq N}\left\{\frac
1{2^{x+1}}\mid{D}_1\mid + \max_{1\leq x \leq N}\mid\eta_y-\xi_y\mid \sum_{l=1}^{x+1}\frac 1{2^{l}}\right\}\nonumber\\
& & \leq \max_{1 \leq x \leq N}\left\{\frac{2^{N-1}}{2^{x+1}}+(1-\frac{1}{2^x})\right\}\max_{1\leq x \leq N}\mid\eta_y-\xi_y\mid\nonumber\\
& & \leq
(1+2^{N-3})\max_{1\leq x \leq N}\mid\eta_y-\xi_y\mid  \nonumber\\
& & \leq (1+2^{N-3})\epsilon_1 =\epsilon_2. \label{calcepsilon}
\end{eqnarray}}
We wish to iterate the above procedure for the next $N-2$
avalanches. We number the avalanches $1,\ldots, N-1$,
and define $\tau_k$ as the time at which the $k$-th avalanche
occurs. As explained for the case $k=1$, we choose all additions
equal, except at times $\tau_k$, where we choose
$\hat{U}^{\xi}(\tau_k)-\hat{U}^{\eta}(\tau_k) = D_k$, with
$$
D_k=\sum_{y=1}^{N-k}2^{y-1}[\hat{\eta}_y(\tau_{k-1})-\hat{\xi}_y(\tau_{k-1})]
$$
and

$$
|D_k| \leq 2^{N-k} \max_{1\leq x \leq
N}|\hat{\eta}_{x}(\tau_{k-1})-\hat{\xi}_{x}(\tau_{k-1})|.
$$
The maximal difference between corresponding sites in the two
resulting configurations has the following bound:
{\setlength\arraycolsep{-27.5pt}
\begin{eqnarray}
\max_{1\leq x \leq N}|\hat{\eta_{x}}(\tau_{k})-\hat{\xi}_{x}(\tau_{k})|\nonumber\\
& & \leq\max_{1 \leq x \leq N}\left\{\frac{1}{2^{x+1}}|D_{k}|+\sum_{l=1}^{x+1}\frac{1}{2^{l}}\max_{1\leq x \leq N}|\hat{\eta}_{x}(\tau_{k-1})-\hat{\xi}_{x}(\tau_{k-1})|\right\}\nonumber\\
& & \leq (1+2^{N-k-2})\max_{1\leq x \leq N}|\hat{\eta}_{x}(\tau_{k-1})-\hat{\xi}_{x}(\tau_{k-1})|\nonumber\\
& & \leq (1+2^{N-k-2})\epsilon_{k} = \epsilon_{k+1}. \nonumber\\
\label{calcepsilon2}
\end{eqnarray}}
Hence for all $k$, $|D_k|$ is bounded from above by
$\epsilon_{k+1} 2^{N-k}$, where $\epsilon_{k+1}$ only depends on
$\epsilon_{k}$ and $N$. With induction, we find,
$$
|D_k| \leq 2^{N-k}\Pi _{l=1}^{k-1}(1+2^{N-l-2})\epsilon_1 := d_k.
$$
We choose $\epsilon_1 = \epsilon_{a,b,N}$ such that $D_{N-1} \leq
\frac {b-a'}4$. As the upper bound $d_{k}$ is increasing in $k$,
we get
 $D_{k} \leq \frac {b-a'}4$, for all $k=1, ..., N-1$.
\hfill $\Box$

\medskip\noindent
{\bf Remark} It follows from our proof that the convergence to the stationary distribution goes in fact exponentially fast. Indeed, every step of the coupling is
such that a certain good event occurs with a certain minimal probability
within a bounded number of steps. Hence, there exists a probabiliy $q >0$ and
a number $M >0$ such that with probability $q$, the coupling is succesfull within
$M$ steps, uniformly in the initial configuration. This implies exponential convergence.  

\section{Zhang's sandpile in infinite volume}
\label{densitysection}
\subsection{Definitions and main results}

In this section we work in general dimension $d$.
We let $\mathcal{X}=[0, \infty)^{\mathbb{Z}^d}$ denote the set of
infinite-volume height configurations in dimension $d$ and
$\Omega=[0, 1)^{\mathbb{Z}^d}$ the set of all stable
configurations. For $x \in \mathbb{Z}^d$, the (Zhang) toppling operator $T_{x}$ 
is defined as
\begin{equation*}
 T_x (\eta)_{y} = \left\{
\begin{array}{ll}
0 & \mbox{ if } x=y,\\
\eta_{y} + \frac{1}{2d}\eta_{x} & \mbox{ if } |y-x|=1,\\
\eta_{y} & \mbox{ otherwise}.
\end{array}
\right.
\end{equation*}

The infinite-volume version of Zhang's sandpile model is quite different
from its abelian sandpile counterpart. Indeed, in the 
infinite-volume abelian sandpile model, it is shown that if a configuration 
can reach a stable one via some order
of topplings, it will reach a stable one by every order of
topplings and the final configuration as well as the topplings
numbers per site are always the same, see \cite{feyredig, quant, anne2}.

In Zhang's sandpile model in infinite volume, the situation is not nearly as nice. Not only does the
final stable realization depend on the order of topplings, the very stabilizability
itself also does. We illustrate this
with some examples.

Consider the initial configuration (in $d=1$)
$$\eta=(\ldots,0, 0, 1.4, 1.2, 0, 0,\ldots).$$
We can reach a stable configuration in any order of the topplings, but the
final configuration as well as the number of topplings per site depend on
which unstable site we topple first. We can choose to start
toppling at the left or right unstable site, or to topple parallel at the same time; 
the different results
are presented in Table \ref{voorbeeldje}.

\begin{table}[ht]
\begin{tabular}{|l|c|c|}
\hline
start at site & toppling numbers & final configuration\\
\hline
left & $(\ldots,0,0,1,1,0,0,\ldots)$ & $(\ldots,0,0.7,0.95,0,0.95,0,\ldots )$ \\
\hline
right & $(\ldots,0,1,2,3,1,0,\ldots)$ &  $(\ldots,0.5,0.5,0.525,0,0.525,0.55 \ldots )$\\
\hline
parallel & $(\ldots,0,0,1,1,0,0,\ldots)$&  $(\ldots,0,0.7,0.6,0.7,0.6,0,\ldots )$ \\
\hline
\end{tabular}
\caption{The three possible stabilizations of $(\ldots,0, 0, 1.4,
1.2, 0, 0,\ldots)$} \label{voorbeeldje}
\end{table}

For a second example, let $$\xi=(\ldots, 0.9, 0.9, 0, 1.4, 1.2, 0, 0.9, 0.9, \ldots).$$
This is a configuration that evolves to a stable configuration in
some orders of topplings, but not by others.
Indeed, if we start to topple the left unstable site first, we obtain
the stable configuration $$(\ldots, 0.9, 0.9, 0.7, 0.95, 0, 0.95, 0.9, 0.9,
\ldots),$$ but if we topple the right unstable site first, after
two topplings, we reach $$\xi'=(\ldots, 0.9, 0.9, 1, 0, 1, 0.6,
0.9, 0.9, \ldots).$$ By arguing as in the proof of the forthcoming Theorem
\ref{voorbeelden}), one can see that this configuration cannot evolve to a stable
configuration.

In view of these examples, we have to be more precise about the way we
perform topplings. In the present paper, we will use the {\em Markov toppling process}: to each site we
associate an independent rate 1 Poisson process. When the
Poisson clock `rings' at site $x$ and $x$ is unstable at that moment, we
perform a Zhang-toppling at that site. If $x$ is stable, we do nothing. 
By $\eta(t)$, we denote the
random configuration at time $t$. We denote by $M(x, t, \eta)$ the (random) number of
topplings at site $x$ up to and including time $t$.

\begin{definition}
A configuration $\eta\in \mathcal{X}$ is said to be
{\em stabilizable} if for every $x\in\mathbb{Z}^d$, 
$$
\lim_{t \to \infty} M(x, t, \eta) < \infty
$$
a.s. In that case we denote the limit random
variable by $M(x, \infty, \eta)$.
\label{stabilizable}
\end{definition}

Denote the collection of all stabilizable configurations by
$\mathcal{S}$. It is not hard to see that $\mathcal{S}$ is shift-invariant and
measurable with respect to the usual Borel sigma field. Hence, if $\mu$ is an ergodic
stationary probability measure on $\mathcal{X}$, $\mu(S)$ is either 0 or 1.
 
\begin{definition}
A probability measure $\mu$ on $\mathcal{X}$ is called 
{\em stabilizable} if $\mu(S)=1$.
\end{definition}

Here is our main result.

\begin{theorem}
Let $\mu$ be an ergodic translation-invariant probability measure on
$\mathcal{X}$ with $\mathbb{E}_{\mu}(\eta_0)=\rho<\infty$. Then
\begin{enumerate}
\item If $\rho \geq 1$, then $\mu$ is not stabilizable, that is, $\mu(S)=0$. 
\item If $0\leq
\rho < \frac{1}{2}$, then $\mu$ is stabilizable, that is, $\mu(S)=1$.
\end{enumerate}
\label{main}
\end{theorem}

The situation when $\frac{1}{2} \leq \rho  < 1$ is not nearly as elegant. Clearly,
when we take $\mu_{\rho}$ to be the point mass at the configuration with constant height 
$\rho$, then $\mu_{\rho}$ is stabilizable for all $\frac{1}{2} \leq \rho < 1$. On the other hand, we have the following theorem, showing that there are measures $\mu$ with
density close to $\frac{1}{2}$ and close to 1 which are not stabilizable.

\begin{theorem}
\label{voorbeelden}
For all $1/2 \leq \rho  < d/(2d-1)$ and $(2d-1)/(2d) < \rho <1$, there exists an ergodic measure $\mu_{\rho}$ with density $\rho$ which is not stabilizable.
\end{theorem}

\medskip\noindent
{\bf Remark} There is no obvious monotonicity in the density as far as 
stabilizability is concerned. Hence we cannot conclude from the previous theorem
that for {\em all} $ 1/2 \leq \rho < 1$ there exists an ergodic measure $\mu_{\rho}$ which is not stabilizable.

\section{Proofs for the infinite-volume sandpile}
Starting from configuration $\eta=(\eta_x)$, we define the total amount of mass that is lost
from site $x$ via topplings, until and including time $t$, by $L(x, t,\eta)$. We
have the following identity.

\begin{equation}
\eta_x(t)=\eta_{x} - L(x, t, \eta) + \frac{1}{2d}\sum_{\mid y-x \mid =1}L(y, t,
\eta). \label{expression}
\end{equation}

The following lemma is easy to understand, and we give it without proof. We write
$\eta(\infty)$ for the (random) limiting configuration when we start with $\eta$.

\begin{lemma}
For a stabilizable $\eta\in\mathcal{X}$, we have
\begin{enumerate}
\item For every $x\in\mathbb{Z}^d$, there exists a random variable $L(x, \infty, \eta) < \infty$ such that a.s.
$$
L(x, t, \eta) \uparrow L(x, \infty, \eta),
$$ 
as $t\rightarrow\infty$. 
\item The following identity holds:
\begin{equation*}\eta(\infty)=\eta(0)+\Delta L(\cdot, \infty, \eta),
\end{equation*}
with $\Delta$ the matrix defined by
$$
\Delta_{x,y} = \left\{
\begin{array}{ll}
-1& \mbox{ if } x=y,\\
\frac{1}{2d}& \mbox{ if } |y-x|=1,\\
0 & \mbox{ otherwise}.
\end{array}
\right.
$$ 
\end{enumerate}
\label{limit}
\end{lemma}

For an initial measure $\mu$, $\mathbb{E}_\mu$ and $\mathbb{P}_{\mu}$
denote the corresponding expectation and probability measures in
the stabilization process. We first show that no mass is lost in the toppling process.
\begin{proposition}
Let $\mu$ be an ergodic shift-invariant probability measure on
$\mathcal{X}$ with
\begin{equation*} \mathbb{E}_{\mu}(\eta_0)=\rho<\infty
\end{equation*}
\noindent which evolves according to the Markov toppling process. Then we have
\begin{enumerate}
\item for $0\leq t <\infty$, $\mathbb{E}_{\mu}\eta_{0}(t)=\rho$,
\item if $\mu$ is stabilizable, then
$\mathbb{E}_{\mu}\eta_{0}(\infty)=\rho$.
\end{enumerate}
\label{keepdensity}
\end{proposition}

\begin{proof} We prove 1.\ via the well known mass transport principle. Let the
initial configuration be denoted by $\eta$. Imagine that
at time $t=0$ we have a certain amount of mass at each site, and we colour all mass white,
except the mass at a special site $x$ which we colour black. Whenever a site topples, we further 
imagine that the black and white mass present at that site, are {\em both} equally distributed
among the neighbours. So, for instance, when $x$ topples for the first time, all its neighbours
receive a fraction $1/(2d)$ of the original black mass at $x$, plus possibly some white mass.
We denote by $B(y,t)$ the total amount of black mass at site $y$ at time $t$. 
First, we argue that at any finite time $t$, 
\begin{equation}
\label{eva}
\sum_{y \in \mathbb{Z}^d}B(y,t)= \eta_x,
\end{equation}
that is, no mass is lost at any finite time $t$. Indeed, had this not been the case, then we define
$t^*$ to be the infimum over those times $t$ for which (\ref{eva}) is not true. Since mass must
be subsequently passed from one site to the next, this implies that there is a path 
$(x=x_0, x_1,\ldots)$ of neighbouring sites to infinity, starting at $x$ such that the sites $x_i$ all topple before
time $t^*$, in the order given. Moreover, since $t^*$ is the infimum, the toppling times $t_i$ of
$x_i$ satisfies $\lim_{i \to \infty} t_i =t^*$. Hence, for any $\epsilon >0$, we can find $i_0$ so
large that for all $i > i_0$, $t_i > t^*-\epsilon.$ Call a site open of its Poisson clock `rings' in
the time interval $(t^*-\epsilon, t^*)$, and closed otherwise. This constitutes an independent percolation process, and if $\epsilon$ is sufficiently small, the open sites do not percolate. Hence
a path as above cannot exist, and we have reached a contradiction. It follows that no mass is lost
at any finite time $t$, and we can now proceed to the routine proof of 1.\ via mass-transport. 

We denote by $X(x, y, t, \eta)$ the amount of mass at $y$ at time $t$ which started at $x$. From
mass preservation, we have
\begin{equation}
\eta_x=\sum_{y\in\mathbb{Z}^d}X(x, y, t, \eta)
\end{equation}
and
\begin{equation}
\eta_y(t)=\sum_{x\in\mathbb{Z}^d}X(x, y, t, \eta).
\end{equation}
Since all terms are non-negative and by symmetry, this gives
\begin{eqnarray*}
\mathbb{E}_{\mu}\eta_0(0)&=&\sum_{y\in\mathbb{Z}^d}\mathbb{E}_{\mu}X(0, y, t,\eta)\\
&=& \sum_{y\in\mathbb{Z}^d}\mathbb{E}_{\mu}X(y, 0, t, \eta) = \mathbb{E}_{\mu}\eta_0(t).
\end{eqnarray*}

To prove 2., we argue as follows. From 1.\ we have that for every $t<\infty$,
$\mathbb{E}_\mu\eta_0(t)=\rho$. Using Fatou's lemma we obtain
\begin{equation}
\mathbb{E}_\mu(\eta_0(\infty))=\mathbb{E}_\mu(\lim_{t\rightarrow\infty}\eta_0(t))\leq\liminf_{t\rightarrow\infty}\mathbb{E}_{\mu}(\eta_0(t))=\rho,
\end{equation}
and therefore it remains to show that
$\mathbb{E}_\mu(\eta_0(\infty))\geq\rho$. We do this by letting a random walk run,
in the same spirit as in \cite{anne2}.

Let $X_{n}, n=0,1,\ldots$ be a simple random walk starting at the origin. We write 
$\mathbb{E}_{rw}, \mathbb{P}_{rw}$ for the expectation and
probability with respect to this random walk. Let $\eta$ be
stabilizable, and let $\eta$ stabilize to a (random realisation) $\eta(\infty)$. The numbers $L(x, \infty, \eta)$ are then also fixed. For all sites $x$ we have the identity
$$
\eta_x(\infty)=\eta(x)-L(x, \infty, \eta)+\frac{1}{2d}\sum_{\mid y-x \mid=1}L(y,
\infty, \eta).
$$
Plugging in $x =X_k$, taking expectations and observing that
$$
\mathbb{E}_{rw}\left(\frac{1}{2d}\sum_{\mid y-X_k \mid=1}L(y,
\infty, \eta)\right) = \mathbb{E}_{rw}(L(X_{k+1}, \infty, \eta)),
$$
we obtain
$$
\mathbb{E}_{rw}\left(\eta_{X_k}(\infty)-\eta_{X_k}\right) =
\mathbb{E}_{rw}(L(X_{k+1}, \infty, \eta)) - \mathbb{E}_{rw}(L(X_{k}, \infty, \eta)).
$$
Now sum from $k=0$ to $k=n-1$ and divide by $n$ to obtain
$$
\frac{1}{n}\mathbb{E}_{rw}\left(\sum_{k=0}^{n-1}\eta_{X_k}(\infty)-\eta_{X_k}\right)
=\frac{1}{n}\mathbb{E}_{rw}\left(L(X_n,\infty,\eta)-L(0,\infty, \eta)\right).
$$
Since $L(0,\infty, \eta)$ is a fixed number, letting $n \to \infty$ gives
\begin{equation}
\liminf_{n\rightarrow\infty}\frac{1}{n}\mathbb{E}_{rw}(\sum_{k=0}^{n-1}\eta_{X_k}(\infty))\geq\limsup_{n\rightarrow\infty}\frac{1}{n}\mathbb{E}_{rw}(\sum_{k=0}^{n-1}\eta_{X_k})
\label{randomwalk2}
\end{equation}
Now, we choose $\eta$ according to $\mu$. Since $\mu$ is
ergodic, the law of $\eta(\infty)$ is also ergodic. It is well known that the scenery processes $\{\eta_{X_n}\}$ and $\{\eta_{X_n}(\infty)\}$ are then also ergodic. Hence, the liminf and
limsup are in fact limits; the right hand of \eqref{randomwalk2} is equal to $\rho$, and the left hand is equal to $\mathbb{E}_\mu(\eta_0(\infty))$.
\end{proof}
 
\begin{proposition}
Let $\eta(t)$ be obtained by the Markov toppling process starting
from $\eta \in \mathcal{X}$. Let $\Lambda$ be a finite subset of
$\mathbb{Z}^d$, such that all sites in $\Lambda$ toppled at least
once before time $t$. Let $\beta_\Lambda$ be the number of
internal bonds of $\Lambda$. Then
\begin{equation}
\sum_{x \in \Lambda} \eta_x(t) \geq \frac {1}{2d} \beta_\Lambda.
\label{minamount}
\end{equation}
\label{interbound}
\end{proposition}

\begin{proof}
Let $(x,y)$ be an internal bond of $\Lambda$. By assumption, both $x$ and $y$
topple before time $t$. Suppose that $x$ is the last to topple among $x$ and $y$.
As a result of this toppling, at least mass $1/(2d)$ is transferred from $x$ to $y$
and this mass will stay at $y$ until time $t$ since $y$ does not topple anymore
before time $t$. In this way, we associate with each internal bond, an amount of
mass of at least $1/(2d)$, which is present in $\Lambda$ at time $t$. Hence
the total amount of mass in $\Lambda$ at time $t$ is at least $1/(2d)$ times
the number of internal bonds.
\end{proof}

We can now prove Theorem \ref{main}.

\medskip\noindent
{\em Proof of Theorem \ref{main}.}
We prove 1.\ first. Let $\mu$ be any ergodic
shift-invariant measure with $\mathbb{E}_{\mu}(\eta_0)=\rho\geq
1$ and suppose $\mu$ is stabilizable. According to Proposition
\ref{keepdensity}, we have 
\begin{equation}
\mathbb{E}_{\mu}(\eta_{0}(\infty))=\mathbb{E}_{\mu}(\eta_0)=\rho\geq1,
\end{equation}
which contradicts the assumption that $\eta(\infty)$ is stable.

For 2., let $\mu$ be any ergodic
shift-invariant probability measure on $\mathcal{X}$ with
$\mathbb{E}_{\mu}(\eta_0)=\rho<\frac{1}{2}$, and suppose that $\mu$ is not
stabilizable. We will now show that
this leads to a contradiction.

Define $C_{n}(t)$ to be the event that before time $t$, every site in
the box $[-n, n]^d$ topples at least once.
Since $\mu$ is not stabilizable we have that
$\mathbb{P}_{\mu}(C_{n}(t))\rightarrow 1$ as $t\rightarrow\infty$.
Indeed, if a configuration is not stabilizable, all sites will topple infinitely many
times as can be easily seen.

Choose $\epsilon>0$ such that $1-\epsilon>2\rho$. For this
$\epsilon$, there exists a non-random time $T^{\epsilon}>0$ such
that for all $ t > T^{\epsilon}$,
\begin{equation}
\mathbb{P}_{\mu}(C_{n}(t))>1-\epsilon.
\end{equation}
From Proposition \ref{interbound} we have that at time $t\geq
T^{\epsilon}>0$, with probability at least $1-\epsilon$, the
following inequality holds:
\begin{equation}
\frac{\sum_{x\in [-n,
n]^d}\eta_{x}(t)}{(2n+1)^d}\geq\frac{1}{2}\frac{(2n)^d}{(2n+1)^d}.
\end{equation}
Therefore, we also have
\begin{equation*}\mathbb{E}_{\mu}\left(\frac{\sum_{x\in [-n, n]^d}\eta_{x}(t)}{(2n+1)^d}\right)
\geq\frac{1}{2}(1-\epsilon)\frac{(2n)^d}{(2n+1)^d}.
\end{equation*}
Since $2\rho<1$, we can choose $n$ so large that
\begin{equation*}(1-\epsilon)\frac{(2n)^d}{(2n+1)^d}>2\rho.
\end{equation*}
Using the shift-invariance of $\mu$ and the toppling
process, for  $t\geq{T}^{\epsilon}$, we find
\begin{equation}
\mathbb{E}_{\mu}\eta_{0}(t)=\mathbb{E}_{\mu}\left(\frac{\sum_{x\in [-n,
n]^d}\eta_{x}(t)}{(2n+1)^d}\right)\geq\frac{1}{2}(1-\epsilon)\frac{(2n)^d}{(2n+1)^d}>\rho.
\end{equation}
However, from Proposition \ref{keepdensity}, we have for any finite
$t$ that $\mathbb{E}_{\mu}\eta_{0}(t)=\mathbb{E}_{\mu}\eta_0(0)=\rho$.
\qed

\medskip\noindent
{\em Proof of Theorem \ref{voorbeelden}.}
We start with $\rho > (2d-1)/(2d)$.
To understand the idea of the argument, it is useful to first assume that we 
have an unstable configuration $\eta$ on a {\em bounded} domain $\Lambda$ (with
periodic boundary conditions) with the property that $\eta_x \geq 1-1/(2d)$, for all $x \in \Lambda$. On such a bounded
domain, we can order the topplings according to the time at which they occur. Hence we can find a sequence of sites $x_1, x_2,\ldots$ (not necessarily all distinct) and a sequence of times
$t_1 < t_2 < \cdots$ such that the $i$-th toppling takes place at site $x_i \in \Lambda$ at time $t_i$. At time $t_1$, $x_1$ topples, so all neighbours of $x_1$
receive at least $1/(2d)$ from $x_1$. This means that all neighbours of $x_1$ become unstable at time $t_1$, and therefore they will all topple at some moment in the future. As a result, $x_1$ itself will also again be unstable after all its neighbours have toppled, and hence $x_1$ will topple again in the future.

In an inductive fashion, assume that after the $k$-th toppling (at site $x_k$ at time
$t_k$), we have that it is certain that all sites that have toppled so far, will
topple again in the future, that is, after time $t_k$. Now consider the next toppling,
at site $x_{k+1}$ at time $t_{k+1}$. If none of the neighbours of $x_{k+1}$ have toppled before, then a similar argument as for $x_1$ tells us that $x_{k+1}$ will topple again in the future. If one or more neighbours of $x_{k+1}$ have toppled before, then the inductive hypothesis implies that they will topple again after time $t_{k+1}$. Hence, we conclude that {\em all} neighbours of $x_{k+1}$ will topple again which implies, just as before, that $x_{k+1}$ itself will topple again.
We conclude that each sites which topples, will topple again in the future, and therefore the configuration can not be stabilized. 

This argument used the fact that we work on a bounded domain, since only then is there
a well-defined sequence of consecutive topplings. But with some extra work, we can make
a similar argument work for the infinite-volume model as well, as follows.

Let $s_0>0$ be so small that the probability that the Poisson clock at the
origin `rings' before time $s_0$ is smaller than the critical probability for independent site percolation on the $d$-dimensional integer lattice. Call a site
{\em open} if its Poisson clock rings before time $s_0$. By the choice of $s_0$, all
components of connected open sites are finite. For each such component of open sites, we now order the topplings that took place between time 0 and time $s_0$. For each
of these components, we can argue as in the first paragraphs of this proof, and
we conclude that all sites that toppled before time $s_0$, will topple
again at some time larger than $s_0$. We then repeat this procedure for the
time interval $[s_0, 2s_0]$, $[2s_0, 3s_0], \ldots$, and conclude that at any
time, a site that topples, will topple again in the future. This means that the
configuration is not stabilizable. Hence, if we take a measure $\mu_{\rho}$ such
that with $\mu_{\rho}$-probability 1, all configurations have the properties
we started out with, then $\mu_{\rho}$ is not stabilizable.

Next, we consider the case where $1/2 \leq \rho < d/(2d-1)$. Consider a measure
$\mu_{\rho}$ whose realizations are a.s.\ `checkerboard' patterns in the
following way: any realization is a translation of the configuration  in which 
all sites whose sum of the coordinates is even obtain mass $2\rho$, and all
sites whose sum of coordinates is odd obtain zero mass. Consider a site $x$ with zero 
mass. Since all neighbours of $x$ are unstable, these neighbours will all topple
at some point. By our choice of $\rho$, $x$ will become unstable precisely at the
moment that the {\em last} neighbour topples - this follows from a simple computation.
By an argument pretty much the same as in the first case, we now see
that all sites that originally obtained mass $2\rho$, have the property that
after they have toppled, {\em all} their neighbours will topple again in the future,
making the site unstable again. This will go on forever, and we conclude that the
configuration is not stabilizable. Hence, $\mu_{\rho}$ is not stabilizable.
\qed

\medskip\noindent
{\bf Remark} The argument in case of parallel topplings is simpler and
works for all $\rho \geq 1/2$; in that case
the checkerboard pattern is preserved at all times, preventing stabilization.


\begin{thebibliography}{99}

\bibitem{BTW}
P. Bak, K. Tang and K. Wiesenfeld (1988) Self-organized criticality.
{\it Phys. Rev. A} {\bf 38}, 364-374.

\bibitem{benjamini}
I. Benjamini, R. Lyons, Y. Peres and O. Schramm (1999) Critical
percolation on any nonamenable group has no infinite clusters,
{\em Annals of Probability} {\bf 27}(3), 1347-1356.

\bibitem{dhar}
D. Dhar (1999) Studying self-organized criticality with exactly solved
models, preprint available at http://www.arxiv.org/abs/cond-mat/9909009.

\bibitem{dickman}
{R. Dickman, M. Mu\~{n}oz, A. Vespagnani and S. Zapperi} (2000)
Paths to self-organized criticality, {\em Brazilian Journal of
Physics} {\bf 30}, 27-41.

\bibitem{Feller} W. Feller (1966) {\em An introduction to Probability Theory and
its Applications}, Volume II, Wiley New York.

\bibitem{anne} A. Fey-den Boer, R. Meester, C. Quant, F. Redig (2008)
A probabilistic approach to Zhang's sandpile model, to appear in {\em Communications in
Mathematical Physics}. 

\bibitem{anne2}{A. Fey-den Boer, R. Meester, F. Redig} (2008)
Stabilizability and percolation in the infinite volume sandpile
model, to appear in {\em Annals of Probability}.

\bibitem{feyredig}
A. Fey-den Boer, F. Redig (2005) Organized versus
self-organized criticality in the abelian sandpile model, {\em
Markov Processes and Related Fields} {\bf 11}(3), 425-442.

\bibitem{janosi}
I.M. Janosi(1989) Effect of anisotropy on the self-organized
critical state, {\em Phys. Rev. A} {\bf 42}(2), 769-774. 

\bibitem{quant} R. Meester, C. Quant (2005)
Connections between `self-organized' and `classical' criticality,
{\em Markov Processes and Related Fields} {\bf 11}, 355-370.

\bibitem{meester}
R. Meester, F. Redig and D. Znamenski (2001) The abelian
sandpile model, a mathematical introduction, {\em Markov Proc.
Rel. Fields} {\bf 7}, 509-523.

\bibitem{Norris} J. R. Norris (1997) {\em Markov Chains}, Cambridge Series in
Statistical and Probabilistic Mathematics.

\bibitem{thorisson}
{H. Thorisson} (2000) Coupling, Stationarity, and Regeneration, {\em
Springer Verlag}, New York.

\bibitem{zhang}
{Y.-C. Zhang} (1989) Scaling theory of Self-Organized Criticality,
{\em Phys. Rev. Lett.} {\bf 63}(5), 470-473.

\end{thebibliography}
\end{document}